# Generalized Current-State Opacity With Dynamically Changing Secrets

Dan You
*Department of Electrical and Electronic Engineering*
*University of Cagliari*
Cagliari, Italy
youdan000@hotmail.com

Shouguang Wang
*School of Information & Electronic Engineering*
*Zhejiang Gongshang University*
Hangzhou, China
wsg5000@hotmail.com

Carla Seatzu
*Department of Electrical and Electronic Engineering*
*University of Cagliari*
Cagliari, Italy
carla.seatzu@unica.it

*Abstract*—**Opacity, an information-flow property related to the privacy and security of a system, has been extensively studied in the context of discrete event systems. Although various notions of opacity have been proposed, in all cases the considered secret was constant. This work focuses on current-state opacity, considering a scenario where the secret changes dynamically with the system evolution. In other words, we propose the new notion of generalized current-state opacity (GCSO), which is with respect to a dynamic-secret model rather than a constant secret. Moreover, we provide a method to verify GCSO based on the construction of the GCSO-verifier. Finally, a practical example is given to illustrate the proposed notion and the method for its verification.**

## I. Introduction

This work investigates an information-flow property called *opacity*, which is related to the privacy and security of a system. In simple words, a system is said to be *opaque* if its secret can never be revealed to an external observer (i.e., intruder) who typically has the full knowledge of the system model but limited observability on the system behavior. Opacity is initially proposed in computer science for the analysis of cryptographic protocols [4] and later extended to discrete event systems (DESs) in the framework of Petri nets [2], labeled transition systems [1], and finite-state automata [5]. Since then, considerable attention has been drawn on opacity in the DES community. By considering different security requirements and different real scenarios, a variety of notions of opacity have been proposed such as *current-state opacity*, *initial-state opacity*, *K-step opacity* and *infinite-step opacity*. Focusing on a specific notion of opacity, the verification and enforcement of opacity are two problems investigated in most existing works [3].

To our best knowledge, although there are a variety of notions of opacity in the literature, all of them consider a constant secret. However, in some practical scenarios, it could happen that the secret of a system changes with the system evolution. For example, in a system modelling the routine activities of a person, his/her opponent may be interested in detecting different secrets of the person before and after observing a particular activity of him/her. Also, some critical information of a system might change due to the occurrence of a particular event that is even unobservable to its intruder. Consequently, we are motivated to generalize the notion of opacity so that the new notion may characterize the privacy/security of a system in the case that the secret dynamically changes with the system evolution.

In this manuscript, we focus on current-state opacity and consider systems modelled by finite-state automata. First, we define a *dynamic-secret model* that describes the changing of the secret with the system evolution. Then, we propose a new notion called *generalized current-state opacity* (GCSO) that corresponds to opacity with respect to a dynamic-secret model rather than a constant secret. A system is generalized current-state opaque (GCSO) if a malicious intruder that observes the system evolution, can never know with certainty that the current state of the system is in the current secret. In addition, we construct a so-called *GCSO-verifier*, based on which we may verify if a system is GCSO. Finally, a case study is provided regarding a moving vehicle that performs some loading and unloading operations, assuming that the intruder is interested in detecting different secret positions of the vehicle depending on whether or not the vehicle has been loaded.

## II. Preliminaries

Let $\Sigma$ be an alphabet and $\Sigma^*$ be the set of all finite-length strings of elements in $\Sigma$ including the empty string $\epsilon$. Given a string $\sigma \in \Sigma^*$, we use $|\sigma|$ to denote the *length* of $\sigma$ ($|\epsilon|=0$).

In this paper, discrete event systems (DES) are modeled as *deterministic finite-state automata* (DFA). A DFA is a quadruple
$$G=(X, \Sigma, \delta, x_0),$$
where $X$ is the finite set of *states*, $\Sigma$ is the set of *events*, $\delta: X \times \Sigma \to X$ is the deterministic *transition function*, and $x_0 \in X$ is the unique *initial state*. The transition function $\delta$ is also extended to the domain $X \times \Sigma^*$ recursively, namely, $\forall x \in X$, $\delta(x, \epsilon)=x$ and $\delta(x, uv)= \delta(\delta(x, u), v)$, $\forall u \in \Sigma^*$, $v \in \Sigma$. The set $L(G)=\{\sigma \in \Sigma^* | \ \delta(x_0, \sigma)!\}$ is the *language* generated by $G$, where "!" means "is defined".

The event set $\Sigma$ of $G$ is partitioned into two disjoint sets, $\Sigma_o$ the set of *observable* events and $\Sigma_u$ the set of *unobservable* events, i.e., $\Sigma=\Sigma_o \cup \Sigma_u$ and $\Sigma_o \cap \Sigma_u=\varnothing$. The *natural projection* is

$P: \Sigma^* \to \Sigma_o^*$ such that $P(\epsilon) = \epsilon$ and

$$\forall u \in \Sigma^*, v \in \Sigma, \; P(uv) = \begin{cases} P(u)v & \text{if } v \in \Sigma_o, \\ P(u) & \text{otherwise.} \end{cases}$$

The natural projection is also extended to the domain $2^{\Sigma^*}$ such that $\forall L \in 2^{\Sigma^*}$, $P(L) = \{P(\sigma) | \sigma \in L\}$. Moreover, given a DFA $G$, we define $P_G^{-1}: \Sigma_o^* \to L(G)$ such that $P_G^{-1}(\omega) = P^{-1}(\omega) \cap L(G)$, i.e., $P_G^{-1}(\omega) = \{\sigma \in L(G) | P(\sigma) = \omega\}$. We call $\sigma \in P_G^{-1}(\omega)$ a *consistent string* of $\omega$ in $G$.

## III. Generalized Current-State Opacity

In this section, we first review the existing notion of current-state opacity with respect to a secret defined as a set of states. Then, we generalize the notion of current-state opacity by taking into account a dynamically changing secret.

*Definition* 1 (*Current-State Opacity*) [5]: Given a system $G = (X, \Sigma, \delta, x_0)$, a secret $S \subseteq X$, and a set of observable events $\Sigma_o \subseteq \Sigma$, system $G$ is *current-state opaque* with respect to $S$ and $\Sigma_o$ if

$(\forall \sigma \in L(G): \delta(x_0, \sigma) \in S)$
$[\exists \sigma' \in L(G): P(\sigma) = P(\sigma') \text{ and } \delta(x_0, \sigma') \notin S]$.

Now, we generalize the notion of current-state opacity considering a secret that dynamically changes with the system evolution. To this aim, we define a model to describe the changing of the secret, which is formalized as follows and is called *dynamic-secret model*.

*Definition* 2 (*Dynamic-secret Model*): Given a system $G = (X, \Sigma, \delta, x_0)$, a *dynamic-secret model* relative to $G$ is a DFA $H = (Y, \Sigma_Y, \delta_Y, y_0)$ such that

1) $Y \subseteq 2^X$;
2) $\Sigma_Y = \Sigma$; and
3) $\forall y \in Y, \forall t \in \Sigma_Y, \delta_Y(y, t)!$.

In words, each state of a dynamic-secret model $H$ is a subset of states of system $G$, representing a secret. Moreover, a dynamic-secret model $H$ is a complete DFA with the same set of events as system $G$. Thus, it follows that $L(G) \subseteq L(H)$ and $H$ actually provides the complete information whether the secret changes due to the occurrence of any event of system $G$.

*Example* 1: Consider the system $G = (X, \Sigma, \delta, x_0)$ in Fig. 1 (a). The model $H = (Y, \Sigma_Y, \delta_Y, y_0)$ in Fig. 1 (b) is a dynamic-secret model relative to $G$. According to $H$, we can see that the initial secret of the system $G$ is $S_1 = \{A, B\}$. Then, every time the event $u_2$ occurs in the system $G$, the secret changes. Specifically, the secret switches between $S_1 = \{A, B\}$ and $S_2 = \{B, C\}$. The occurrence of other events does not change the current secret. ∎

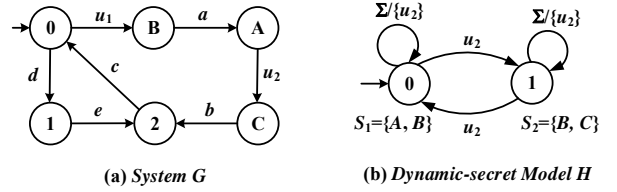

Fig. 1 System $G$ and its relative dynamic-secret model $H$

In what follows, we propose the notion of *generalized current-state opacity* (GCSO) with respect to a dynamic-secret model.

*Definition* 3 (*Generalized Current-State Opacity*): Given a system $G = (X, \Sigma, \delta, x_0)$, a dynamic-secret model $H = (Y, \Sigma_Y, \delta_Y, y_0)$ relative to $G$, and a set of observable events $\Sigma_o \subseteq \Sigma$, system $G$ is *generalized current-state opaque* (GCSO) with respect to $H$ and $\Sigma_o$ if

$(\forall \sigma \in L(G): \delta(x_0, \sigma) \in \delta_Y(y_0, \sigma))$
$[\exists \sigma' \in L(G): P(\sigma) = P(\sigma') \text{ and } \delta(x_0, \sigma') \notin \delta_Y(y_0, \sigma')]$.

In simple words, a system $G$ is GCSO with respect to $H$ and $\Sigma_o$ if for any string $\sigma$ that is generated by $G$ and that leads to a state belonging to the corresponding secret, which is uniquely identified generating $\sigma$ on $H$, there exists another string $\sigma'$ that produces the same observation as $\sigma$ and whose corresponding state does not belong to the corresponding secret. It is worth noting that the secret of the system may change with the system evolution. Thus, strings $\sigma$ and $\sigma'$ may determine two different secrets, i.e., it may occur that $\delta_Y(y_0, \sigma) \neq \delta_Y(y_0, \sigma')$. Consequently, the physical meaning of a GCSO system is that its intruder can never know with certainty that the current state of the system is in its current secret. On the contrary, if the system is not GCSO, then there exists at least one observation based on which the intruder knows with certainty that the current state of the system is in its current secret. Note that we are assuming that the intruder is not interested in knowing which the current secret is.

*Example* 2: Consider again the system $G = (X, \Sigma, \delta, x_0)$ and the dynamic-secret model $H = (Y, \Sigma_Y, \delta_Y, y_0)$ in Fig. 1. Suppose that $\Sigma_o = \{a, b, c, d, e\}$ and $\Sigma_u = \{u_1, u_2\}$. We want to show that $G$ is not GCSO with respect to $H$ and $\Sigma_o$. Consider the string $\sigma_1 = u_1 a$ that leads to state A in $G$ and corresponds to the secret $S_1 = \{A, B\}$ on $H$, and thus it is $A \in S_1$. We can see that the same observation of $\sigma_1$, namely $a$, can be produced by only another string, i.e., $\sigma_2 = u_1 a u_2$. String $\sigma_2$ leads to state C in $G$ and identifies the secret $S_2 = \{B, C\}$ on $H$. Therefore, it is $C \in S_2$. Thus, according to Definition 3, system $G$ is not GCSO with respect to $H$ and $\Sigma_o$. ∎

The following proposition presents a necessary and sufficient condition for GCSO.

*Proposition* 1: Given a system $G = (X, \Sigma, \delta, x_0)$, a dynamic-secret model $H = (Y, \Sigma_Y, \delta_Y, y_0)$ relative to $G$, and a set of observable events $\Sigma_o \subseteq \Sigma$, system $G$ is GCSO with respect to $H$ and $\Sigma_o$ iff

$\forall \omega \in P(L(G)), \exists \sigma \in P_G^{-1}(\omega): \delta(x_0, \sigma) \notin \delta_Y(y_0, \sigma)$.

*Proof*: (=>) Straightforward from Definition 3.
(<=) By contradiction, suppose that
$\exists \omega \in P(L(G)), \forall \sigma \in P_G^{-1}(\omega), \delta(x_0, \sigma) \in \delta_Y(y_0, \sigma)$.
Let $\sigma_1 \in P_G^{-1}(\omega)$. It thus follows that $\delta(x_0, \sigma_1) \in \delta_Y(y_0, \sigma_1)$. Since system $G$ is GCSO with respect to $H$ and $\Sigma_o$, by Definition 3, $\exists \sigma_2 \in P_G^{-1}(\omega)$, s.t. $\delta(x_0, \sigma_2) \notin \delta_Y(y_0, \sigma_2)$, which however contradicts the fact that $\forall \sigma \in P_G^{-1}(\omega), \delta(x_0, \sigma) \in \delta_Y(y_0, \sigma)$. Consequently, it holds that
$\forall \omega \in P(L(G)), \exists \sigma \in P_G^{-1}(\omega), \delta(x_0, \sigma) \notin \delta_Y(y_0, \sigma)$. ∎

*Example* 3: Consider again the system $G=(X, \Sigma, \delta, x_0)$ and the dynamic-secret model $H=(Y, \Sigma_Y, \delta_Y, y_0)$ in Fig. 1, still assuming that $\Sigma_o=\{a, b, c, d, e\}$ and $\Sigma_u=\{u_1, u_2\}$. Since there exists an observation $a$ with $P_G^{-1}(a)=\{u_1a, u_1au_2\}$ such that $\delta(x_0, u_1a) \in \delta_Y(y_0, u_1a)$ and $\delta(x_0, u_1au_2) \in \delta_Y(y_0, u_1au_2)$, by Proposition 1 system $G$ is not GCSO with respect to $H$ and $\Sigma_o$. ∎

*Remark* 1: In our problem setting, every evolution (i.e., every string) identifies a pair (current state, current secret). To see if a system is GCSO, we should consider states and secrets pairs by pairs. For example, consider an observation $\omega$ that is produced only by two evolutions $\sigma_1$ and $\sigma_2$. Assume that $\sigma_1$ determines the state $x_1$ and the secret $S_1$ and $\sigma_2$ determines the state $x_2$ and the secret $S_2$. Therefore, we should check if the state $x_1$ is in the secret $S_1$ and if the state $x_2$ is in the secret $S_2$. Several cases could occur. Some of them are sketched in Fig. 2. (Note that not all the cases are enumerated due to the space constraint). If we ignore other observations but only consider the observation $\omega$, the system is not GCSO in Cases (a)-(c) while it is GCSO in Cases (d)-(f). In particular, looking at Case (d), although all the possible current states are included in the union of the possible secrets, the system is actually GCSO. This is because the state $x_1$ is not in the secret $S_1$ if we consider the evolution $\sigma_1$ (or the state $x_2$ is not in the secret $S_2$ if we consider the evolution $\sigma_2$). ∎

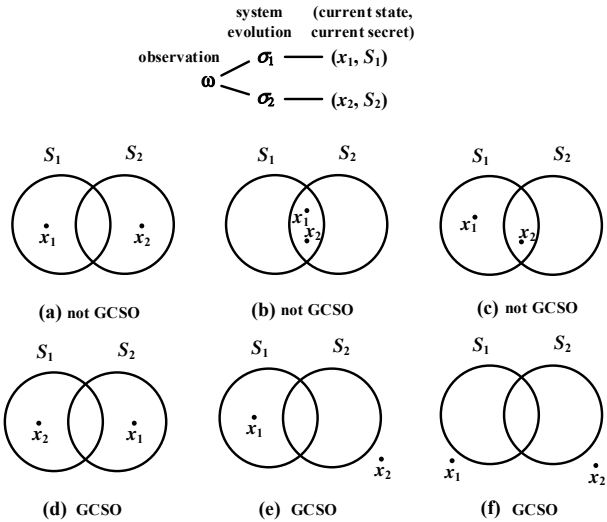

Fig. 2 A series of possible cases that could appear when observing $\omega$ assuming that $P_G^{-1}(\omega)=\{\sigma_1, \sigma_2\}$

## IV. VERIFICATION OF GENERALIZED CURRENT-STATE OPACITY

In this section, we propose a verification method for GCSO. To this aim, we construct the so-called *GCSO-verifier*.

*Definition* 4: Given a system $G=(X, \Sigma, \delta, x_0)$, a dynamic-secret model $H=(Y, \Sigma_Y, \delta_Y, y_0)$ relative to $G$, and a set of observable events $\Sigma_o \subseteq \Sigma$, the *GCSO-verifier* is defined as a DFA
$$Ver(G, H)=(Z, \Sigma_o, \delta_Z, z_0),$$
where
- $Z \subseteq 2^{X \times Y}$;
- $z_0=\{(\delta(x_0, \sigma), \delta_Y(y_0, \sigma))|\sigma \in P_G^{-1}(\epsilon)\}$;
- $\delta_Z: Z \times \Sigma_o \to Z$ is the transition function such that $\forall z \in Z$, $\forall t \in \Sigma_o$, if $\exists (x, y) \in z$, s.t. $\delta(x, t)!$, $\delta_Z(z, t)$ is defined such that
$$\delta_Z(z, t)=\{(x', y') \in X \times Y \mid \exists (x, y) \in z, \exists \alpha \in \Sigma_u^* \text{ s.t.}$$
$$\delta(x, t\alpha)! \wedge \delta(x, t\alpha)=x' \wedge \delta_Y(y, t\alpha)=y'\}.$$

$Ver(G, H)$ is defined only considering the accessible part from the initial state $z_0$, namely removing all those states of $Ver(G, H)$ that are not reachable from the initial state $z_0$ and all their input and output arcs.

By Definition 4, the GCSO-verifier $Ver(G, H)$ can be constructed as follows. First, we create the initial node $z_0=\{(\delta(x_0, \sigma), \delta_Y(y_0, \sigma))|\sigma \in P_G^{-1}(\epsilon)\}$. Next, for every observable event $t$, if there exists $(x, y) \in z_0$ such that $\delta(x, t)$ is defined, we define $\delta_Z(z_0, t)=\{(x', y') \in X \times Y \mid \exists (x, y) \in z_0, \exists \alpha \in \Sigma_u^*$ s.t. $\delta(x, t\alpha)! \wedge \delta(x, t\alpha)=x' \wedge \delta_Y(y, t\alpha)=y'\}$. If there already exists a node $z$ such that $z=\delta_Z(z_0, t)$, we simply add a transition from node $z_0$ to node $z$ via event $t$; otherwise, we create a new node $z$ such that $z=\delta_Z(z_0, t)$ as well as a transition from node $z_0$ to node $z$ via event $t$ and focusing on the new node $z$, we repeat the above procedure to generate its "son nodes". Finally, the GCSO-verifier is obtained.

We observe that every node of the GCSO-verifier contains a set of pairs whose first entry is a state in system $G$ and whose second entry is a secret identified by dynamic-secret model $H$. For the sake of presentation, we call every such pair a *state-secret pair* in the remainder of the paper.

The following lemma characterizes nodes of the GCSO-verifier and the language generated by the GCSO-verifier.

*Lemma* 1: Given a system $G=(X, \Sigma, \delta, x_0)$, a dynamic-secret model $H=(Y, \Sigma_Y, \delta_Y, y_0)$ relative to $G$, a set of observable events $\Sigma_o \subseteq \Sigma$, and the GCSO-verifier $Ver(G, H)=(Z, \Sigma_o, \delta_Z, z_0)$, it holds that

1) $\forall \omega \in L(Ver(G, H))$,
$$\delta_Z(z_0, \omega)=\{(\delta(x_0, \sigma), \delta_Y(y_0, \sigma)) \mid \sigma \in P_G^{-1}(\omega)\};$$

2) $L(Ver(G, H))=P(L(G))$.

*Proof*: 1) First, consider $\epsilon \in L(Ver(G, H))$. Trivially, $|P_G^{-1}(\epsilon)| \neq \emptyset$. Moreover, it is $\delta_Z(z_0, \epsilon)=z_0=\{(\delta(x_0, \sigma), \delta_Y(y_0, \sigma))|\sigma \in P_G^{-1}(\epsilon)\}$ by Definition 4. Then, consider $\omega'=\omega t \in L(Ver(G, H))$, where $t \in \Sigma_o$ and $\omega \in L(Ver(G, H))$ such that $|P_G^{-1}(\omega)| \neq \emptyset$ and
$$\delta_Z(z_0, \omega)=\{(\delta(x_0, \sigma), \delta_Y(y_0, \sigma)) \mid \sigma \in P_G^{-1}(\omega)\}. \quad (1)$$

By Definition 4,
$$\exists (x, y) \in \delta_Z(z_0, \omega), \text{ s.t. } \delta(x, t)! \quad (2)$$
and
$$\delta_Z(\delta_Z(z_0, \omega), t) = \{(\delta(x, t\alpha), \delta_Y(y, t\alpha)) \mid \exists (x, y) \in \delta_Z(z_0, \omega), \exists \alpha \in \Sigma_u^* \text{ s.t. } \delta(x, t\alpha)!\}. \quad (3)$$
By (1) and (2), we can see that
$$\exists \sigma \in P_G^{-1}(\omega), \text{ s.t. } \delta(x_0, \sigma t)!. \quad (4)$$
By (1) and (3), it follows that
$$\delta_Z(\delta_Z(z_0, \omega), t) = \{(\delta(\delta(x_0, \sigma), t\alpha), \delta_Y(\delta_Y(y_0, \sigma), t\alpha)) \mid \exists \sigma \in P_G^{-1}(\omega), \exists \alpha \in \Sigma_u^* \text{ s.t. } \delta(\delta(x_0, \sigma), t\alpha)!\}.$$
Furthermore, it is
$$\delta_Z(z_0, \omega') = \delta_Z(z_0, \omega t) = \{(\delta(x_0, \sigma t\alpha), \delta_Y(y_0, \sigma t\alpha)) \mid \exists \sigma \in P_G^{-1}(\omega), \exists \alpha \in \Sigma_u^* \text{ s.t. } \delta(x_0, \sigma t\alpha)!\}. \quad (5)$$
Note that
$$P_G^{-1}(\omega') = P_G^{-1}(\omega t) = \{\sigma t\alpha \mid \sigma \in P_G^{-1}(\omega), \alpha \in \Sigma_u^*, \delta(x_0, \sigma t\alpha)!\}. \quad (6)$$
Thus, due to (5) and (6), it is
$$\delta_Z(z_0, \omega') = \{(\delta(x_0, \sigma'), \delta_Y(y_0, \sigma')) \mid \sigma' \in P_G^{-1}(\omega')\}.$$
Also, since $|P_G^{-1}(\omega)| \neq \varnothing$, we have $|P_G^{-1}(\omega')| \neq \varnothing$ by (4) and (6). Finally, we conclude that $\forall \omega \in L(Ver(G, H)), |P_G^{-1}(\omega)| \neq \varnothing$ and $\delta_Z(z_0, \omega) = \{(\delta(x_0, \sigma), \delta_Y(y_0, \sigma)) \mid \sigma \in P_G^{-1}(\omega)\}$.

2) Since $\forall \omega \in L(Ver(G, H)), |P_G^{-1}(\omega)| \neq \varnothing$, it is trivial to see that $\forall \omega \in L(Ver(G, H)), \omega \in P(L(G))$. In other words, $L(Ver(G, H)) \subseteq P(L(G))$.

Now, we prove that $L(Ver(G, H)) \supseteq P(L(G))$. First, consider $\epsilon \in P(L(G))$. Clearly, $\epsilon \in L(Ver(G, H))$. Then, consider $\omega' = \omega t \in P(L(G))$, where $t \in \Sigma_o$ and $\omega \in P(L(G)) \land \omega \in L(Ver(G, H))$. We prove that $\omega' \in L(Ver(G, H))$ as follows. Since $\omega t \in P(L(G))$, $\omega \in P(L(G))$ and $t \in \Sigma_o$, it holds that
$$\exists \sigma \in P_G^{-1}(\omega), \text{ s.t. } \sigma t \in L(G). \quad (7)$$
Since $\omega \in L(Ver(G, H))$, there exists $z \in Z$ such that
$$z = \delta_Z(z_0, \omega) = \{(\delta(x_0, \sigma), \delta_Y(y_0, \sigma)) \mid \sigma \in P_G^{-1}(\omega)\}. \quad (8)$$
By (7) and (8), $\delta_Z(z, t)$ is defined. Thus, $\omega' \in L(Ver(G, H))$. Finally, it is concluded that $\forall \omega \in P(L(G)), \omega \in L(Ver(G, H))$, i.e., $L(Ver(G, H)) \supseteq P(L(G))$.

Consequently, $P(L(G)) = L(Ver(G, H))$. ■

The verification of GCSO can be performed using the GCSO-verifier. Specifically, as indicted by the following theorem, a system is not GCSO if and only if there exists a node in the corresponding GCSO-verifier whose state-secret pairs satisfy the condition that the first entry (i.e., state) belongs to the second entry (i.e., secret).

*Theorem* 1: Given a system $G = (X, \Sigma, \delta, x_0)$, a dynamic-secret model $H = (Y, \Sigma_Y, \delta_Y, y_0)$ relative to $G$, a set of observable events $\Sigma_o \subseteq \Sigma$, and the GCSO-verifier $Ver(G, H) = (Z, \Sigma_o, \delta_Z, z_0)$,

$G$ is not GCSO w.r.t. $H$ and $\Sigma_o$
$\Leftrightarrow \exists z \in Z$, s.t. $\forall (x, y) \in z, x \in y$.

*Proof*: Straightforward from Proposition 1 and Lemma 1. ■

*Example* 4: Consider again the system $G = (X, \Sigma, \delta, x_0)$ and the dynamic-secret model $H = (Y, \Sigma_Y, \delta_Y, y_0)$ in Fig. 1, still assuming that $\Sigma_o = \{a, b, c, d, e\}$ and $\Sigma_u = \{u_1, u_2\}$. The GCSO-verifier $Ver(G, H)$ is depicted in Fig. 3. We can see that there exists a grey node where in each state-secret pair the first entry belongs to the second entry. Thus, $G$ is not GCSO w.r.t. $H$ and $\Sigma_o$ by Theorem 1. ■

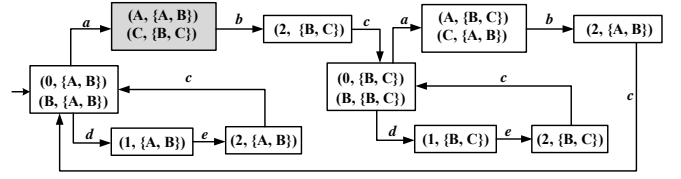

Fig. 3 $Ver(G, H)$ relative to system $G$ and dynamic-secret model $H$ in Fig. 1 with $\Sigma_o = \{a, b, c, d, e\}$ and $\Sigma_u = \{u_1, u_2\}$

*Remark* 2: The GCSO-verifier $Ver(G, H)$ contains at most $2^{X \times 2^X}$ states and $|\Sigma_o| \times 2^{X \times 2^X}$ transitions since $Z \subseteq 2^{X \times Y}$ and $Y \subseteq 2^X$. Thus, the complexity of the verification method is $O(|\Sigma_o| \times 2^{X \times 2^X})$. ■

## V. CASE STUDY

In this section, we provide a practical example where the secret of the considered system dynamically changes.

Consider a vehicle that visits several sites to perform its daily missions, including a delivery mission. In more detail, there are 11 sites, among which two are loading sites (named A and B), one is an unloading site (named C) and the others are passage sites (named 0-7, respectively). The possible movements of the vehicle are modelled by a DFA $G$ as shown in Fig. 4, where each state models the position (namely, the current site) of the vehicle and each transition indicates the movement of the vehicle from one site to another. We assume that sensors are deployed in some areas to detect the movements of the vehicle. Due to physical constraints and/or economic reasons, not all movements are perfectly detectable. Therefore, some movements are unobservable, while some movements generate the same observation as some other movements. Consequently, we consider the DFA model $G$ in Fig. 4, assuming that $\Sigma_o = \{a, b, c, d, e, f, g\}$ and $\Sigma_u = \{u_1, u_2, u_3\}$.

Now, we focus on the delivery mission taken by the vehicle. We assume that it loads product $\alpha$ in loading site A, loads product $\beta$ in loading site B, and unloads products (no matter $\alpha$ or $\beta$) in unloading site C. Moreover, we assume that 1) in the case that the vehicle is empty, it is fully loaded when it arrives at a loading site (A or B); and 2) in the case that the vehicle is loaded, it is fully unloaded when it arrives at the unloading site C.

Suppose that there exists an intruder who wants to know which product the vehicle is delivering. We assume that the intruder has the knowledge of the system model and monitors the movements of the vehicle by partial observations, namely, by the occurrence of observable events.

We consider the following scenario. The intruder detects which product the vehicle is delivering by different means depending on whether or not the vehicle is empty. In the case that the vehicle is empty, the intruder infers with certainty that the vehicle is delivering product $\alpha$ (resp. product $\beta$) if he/she knows for sure that the vehicle is now in site A (resp. site B) since the vehicle is definitely fully loaded by product $\alpha$ (resp. product $\beta$); in the case that the vehicle is loaded, we assume that the intruder can take a particular means in site 2 to see which product the vehicle is delivering when the vehicle visits site 2 and we assume that the intruder is *conservative* in the sense that he/she takes such a means only when he/she knows for sure that the vehicle is now in site 2. As a result, in the case that the vehicle is empty, there are two secrets, namely, {A} and {B}, which should be considered simultaneously, while in the case that the vehicle is loaded, the secret is {2}.

We observe that the intruder may know whether or not the vehicle is empty by observing the occurrence of events $a$ and $b$ in system $G$. Specifically, in the case that the vehicle is empty, if the intruder observes event $a$, he/she knows that the vehicle has been loaded since $a$ is an output event from both loading sites A and B; in the case that the vehicle is loaded, if the intruder observes event $b$, he/she knows that the vehicle is empty again since $b$ is an output event from unloading site C. Consequently, the secret of system $G$ actually dynamically changes with the system evolution, and to be precise, changes with the observation of the intruder in this example. Here, we need the combination of two dynamic-secret models $H_1$ and $H_2$ to describe the changing of the secret in the above scenario, which are depicted in Fig. 5. Note that $H_1$ and $H_2$ focus on the secret {A} and the secret {B}, respectively, in the case that the vehicle is empty. Consider the dynamic-secret model $H_1$. It tells that the secret is $S_1=\{A\}$ at the beginning, i.e., when the vehicle is empty. Once event $a$ is observed, the intruder realizes that the vehicle has already been loaded and thus the secret becomes $S_2=\{2\}$. Then, once event $b$ is observed, the intruder realizes that the vehicle has already been unloaded and thus the secret becomes $S_1=\{A\}$ again. Hence, the secret always switches from $S_1=\{A\}$ to $S_2=\{2\}$ when $a$ is observed and switches from $S_2=\{2\}$ to $S_1=\{A\}$ when $b$ is observed. The dynamic-secret model $H_2$ is the same as $H_1$ except that it is $S_1=\{B\}$ in $H_2$. We consider the combination of $H_1$ and $H_2$ because both the fact that the vehicle is in site A and the fact that the vehicle is in site B should be kept secret to the intruder when the vehicle is empty.

Now, we verify if system $G$ is GCSO w.r.t. $H_1$ and $\Sigma_o$. The GCSO-verifier $Ver(G, H_1)$ is constructed in Fig. 6. Since we cannot find a node in $Ver(G, H_1)$ where each state-secret pair satisfies the condition that the first entry belongs to the second entry, we conclude that system $G$ is GCSO w.r.t. $H_1$ and $\Sigma_o$. Similarly, we conclude that system $G$ is GCSO w.r.t. $H_2$ and $\Sigma_o$ by analyzing the GCSO-verifier $Ver(G, H_2)$, which can be obtained by replacing every element {A} in $Ver(G, H_1)$ with {B} and is thereby omitted due to space constraints. Since system $G$ is both GCSO w.r.t. $H_1$ and $\Sigma_o$ and w.r.t. $H_2$ and $\Sigma_o$, the intruder can never know which product the vehicle is delivering in the considered practical scenario.

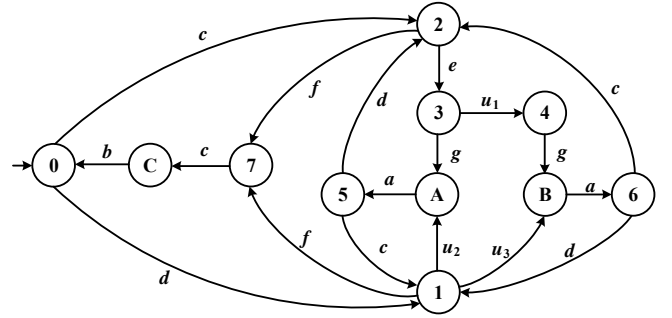

Fig. 4 DFA $G$ modeling the possible movements of a vehicle

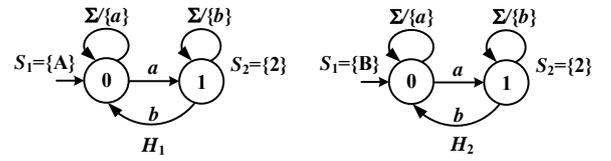

Fig. 5 Dynamic-secret models $H_1$ and $H_2$ relative to system $G$ in Fig. 4

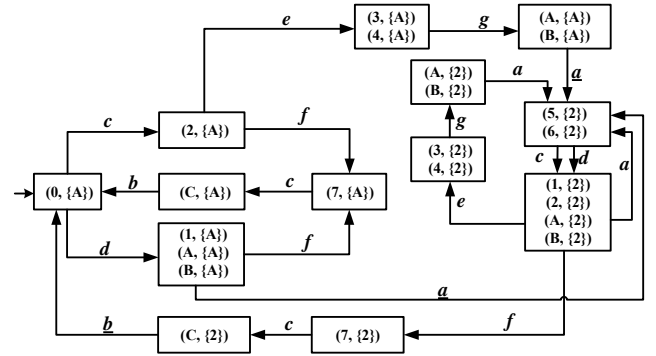

Fig. 6 $Ver(G, H_1)$ relative to system $G$ in Fig. 4 and dynamic-secret model $H_1$ in Fig. 5 with $\Sigma_o=\{a, b, c, d, e, f, g\}$ and $\Sigma_u=\{u_1, u_2, u_3\}$

VI. CONCLUSIONS AND FUTURE WORK

This work generalizes the notion of current-state opacity by taking into account a dynamically changing secret. In particular, a dynamic-secret model is defined that describes the changing of the secret with the evolution of the considered system. Then, we propose the notion of generalized current-state opacity with respect to a dynamic-secret model, which physically characterizes the fact that the intruder can never know with certainty that the current state of a considered system is in the current secret. Moreover, we construct the so-called GCSO-verifier by which we may verify if a system is generalized current-state opaque. Finally, a practical example is provided to illustrate the proposed notion and the verification approach.

As a future work we plan to generalize the notion of a dynamic-secret model so that a single model can used to describe the scenario where multiple secrets are considered simultaneously. Also, we plan to consider the enforcement of the generalized current-state opacity in the case that a system is not generalized current-state opaque.

## REFERENCES

[1] J. W. Bryans, M. Koutny, L. Mazaré, and P. Y. A. Ryan, "Opacity generalised to transition systems," *International Journal of Information Security,* vol. 7, no. 6, pp. 421-435, 2008, doi: 10.1007/s10207-008-0058-x.

[2] J. W. Bryans, M. Koutny, and P. Y. A. Ryan, "Modelling opacity using petri nets," *Electronic Notes in Theoretical Computer Science,* vol. 121, pp. 101-115, 2005.

[3] R. Jacob, J.-J. Lesage, and J.-M. Faure, "Overview of discrete event systems opacity: Models, validation, and quantification," *Annual Reviews in Control,* vol. 41, pp. 135-146, 2016, doi: 10.1016/j.arcontrol.2016.04.015.

[4] L. Mazar´e, "Using unification for opacity properties," presented at the WITS: 4, 2004. [Online]. Available: http://www-verimag.imag.fr/TR/TR- 2004- 24.pdf.

[5] A. Saboori and C. N. Hadjicostis, "Notions of security and opacity in discrete event systems," in *Decision and Control, 2007 46th IEEE Conference on*, 2007, pp. 5056–5061, doi: 10.1109/CDC.2007.4434515.